\documentstyle[aaspp4]{article}
\slugcomment{}
\lefthead{Leitch \& Vasisht}
\righthead{Terrestrial Extinctions}
\begin{document}

\def\msun{$\rm M_\odot$}

\title{Mass Extinctions and The Sun's Encounters with Spiral Arms}

\author{Erik M. Leitch \& Gautam Vasisht}
\affil{California Institute of Technology, 105-24\\Pasadena CA 91125}

\def\simgt{\lower.5ex\hbox{$\; \buildrel > \over \sim \;$}}
\def\simlt{\lower.5ex\hbox{$\; \buildrel < \over \sim \;$}}

\begin{abstract}
{The terrestrial fossil record shows that the exponential rise in
biodiversity since the Precambrian period has been punctuated by large
extinctions, at intervals of $40$ to $140$ Myr. These mass extinctions
represent extremes over a background of smaller events and the natural
process of species extinction. We point out that the non-terrestrial
phenomena proposed to explain these events, such as boloidal impacts
(a candidate for the end-Cretaceous extinction), and nearby
supernovae, are collectively far more effective during the solar
system's traversal of spiral arms.
Using the best
available data on the location and kinematics of the Galactic spiral
structure (including distance scale and kinematic uncertainties),
we present evidence that arm crossings provide a viable explanation 
for the timing of the large extinctions.}
\end{abstract}

The literature is replete with suggestions of non-terrestrial phenomena as 
the candidate causes for large scale extinctions. The most frequently 
invoked are supernovae and boloidal impacts (e.g. comets from 
the Oort Cloud), the latter a strong candidate for the K/T extinction 
ever since the discovery of the Iridium anomaly in the K/T boundary
clay \cite{ALV90}.  Certainly the most violent events in the
solar neighborhood during geologic history would have been supernovae
(barring the possibility of a nearby $\gamma$-ray burst, which is far
less likely; Thorsett 1995);
that the structure of the very local interstellar medium
is considered to be the result of a supernova, possibly related to the
Geminga pulsar (at 150~pc) \cite{BIG96}, is an impressive reminder of
their potential impact.  Supernovae and young supernova remnants,
especially those occurring at distances $\simlt 10$ pc \cite{RUD74},
can result in biospheric imbalance through a variety of processes,
including ozone depletion by enhanced ionizing radiation and cosmic
rays (\cite{SCH95}, \cite{KOY95}), the absorption of visible light by
the formation of NO$_2$ \cite{CRU96}, and in rare cases the direct
deposition of supernova debris.

Tidal and collisional encounters with intermediate-sized gas or dust
clouds might focus cometary activity ($\sim 10^9$ comets) to the inner
solar system, by scattering of Oort Cloud member bodies, a mechanism
proposed to explain the K/T boundary event.  In
addition, the passage of the Sun through a cloud of density $n \simgt
10^4$ cm$^{-3}$ could raise the solar luminosity significantly,
through Bondi accretion, as well as raise the opacity
of Earth's atmosphere, directly affecting the insolation on Earth
\cite{McC75}.  While the recent association of the Chicxulub crater
with the K/T boundary lends credence to the boloidal impact model, the
large concentration of Ir deposited at the boundary may indicate that
accretion also played an important role \cite{YABU}.

The proposed mechanisms constitute a set of plausible external agents
for any one extinction, yet do not of themselves suggest any
explanation for the timing of the mass extinctions, or for
the large variation in severity of the observed extinctions.
  Hatfield
and Camp \cite{HAT81} were among the first to suggest that
extinctions might be correlated with Galactic-plane crossing due to
the solar orbit's vertical oscillations.  Rampino and Stothers (1984),
as well as Schwartz and James \cite{SCH84}, have invoked these
z-oscillations in connection with the suggested $\sim 26 - 30$ Myr
periodicity of minor extinctions, as virtually all of the postulated
extinction mechanisms concentrate toward the Galactic plane.  However,
the fact that we are presently half-way between extinction cycles and
that the Sun's position is nearly midplane implies that Galactic plane
passages are unlikely causes for the extinctions, unless the Sun
has suffered a violent gravitational encounter in the last 15 Myr
(\cite{CLU96}).  Moreover, even if the correlation were exact, it does
not explain the enormous difference in magnitude between the 6 largest
extinctions and extinctions which occur on 30 Myr timescales.

While acknowledging that the apparent quasi-periodicity of mass extinctions
may in fact be spurious, and that the extinction record may be one of
chance encounters of varying magnitude, or indeed merely a record of 
terrestrial cataclysms, we suggest that the spiral arm environment of the 
Galaxy provides a natural framework in which all of the
astrophysical mechanisms discussed thus far would operate most
efficiently.  Pre-supernova stars (the luminous O and B stars) are born
primarily in the spiral arms, and spend much of their short lifetimes
($\simlt2\times10^7$ yr) in their vicinity.  The Type II/Ib
supernovae, which are a consequence of the core-collapse of OB stars,
have a Galactic rate of roughly $R_{S\!N} \approx 1/30$ yr$^{-1}$
\cite{VAN91} and are distributed with a scale height $z \simeq 10^2$ pc in the disk.  The longest lifetime of a pre-supernova star is $\tau
\simeq 2\times 10^7$ yr (for masses $M \simgt 8-9$ \msun).
Defining the effective distance over which a supernova may have a
profound impact on the biosphere as $l_k\simeq 10$~pc (\cite{SCH95}),
the number of significant supernova encounters at the solar Galactic
radius is $N_{S\!N} \simeq R_{S\!N}l_k^3\tau/N_{sp}A_{sp}z$, where
$N_{sp} \simeq 4$ is the number of Galactic arms. The influence area
$A_{sp}$ is assumed to be roughly the product of the arm-length $\sim
10$ kpc and the arm-width, i.e. $(\Omega_\odot-\Omega_p)R_0\tau$, where
$\Omega_\odot$ is the angular speed at the solar galactocentric radius, $\Omega_p$ is the pattern speed of the spiral arms (see below) and $\tau \sim 10^7$ is 
the average lifetime of a supernova
progenitor star.  Then, $$N_{S\!N} \simeq 0.5 \left({R_{SN}\over
0.033~\hbox{yr$^{-1}$}}\right)\left({l\over
10~\hbox{pc}}\right)^3\left({\tau \over
10^{7}~\hbox{yr}}\right)\left({z\over
100~\hbox{pc}}\right)^{-1}\left({A_{sp} \over
6~\hbox{kpc$^2$}}\right)^{-1}\left({N_{sp} \over 4}\right)^{-1}$$ is
the typical number of supernovae encountered within $l_k$ during one
spiral arm passage. In addition, recent X-ray observations have shown
that young supernova remnants (of radius $\simlt$ 10 pc) are active
sites of acceleration of cosmic rays to energies $\simgt 10^2$ TeV
(\cite{KOY95}).  The above estimate shows that the chances of
intercepting a supernova shock front are significant, leading
to sustained exposure ($\sim 10^4$ yr) of the upper atmosphere to
cosmic ray bombardment, by factors $10^2 - 10^3$ over the mean level.

Besides supernovae, gravitational perturbers such as large complexes
of molecular gas and dust (the giant molecular clouds and the
intermediate sized clouds), with typical sizes of a few hundred
parsecs and masses of up to $10^6$~\msun, are also
concentrated along spiral arms. It is instructive, therefore, to trace
the first order solar orbit through the best estimate of the 
structure of the Milky Way and the position of its spiral arms, back to 
the beginning of the Phanerozoic period (0 -- 500 Myr-ago, or
$\simlt 5$\% of the age of the Galactic disk).  Episodes during the solar
motion may then be compared directly with episodes in the geologic
timeline.

In its simplest approximation, the solar revolution is circular with
an adopted galactocentric radius $R_{0}\simeq 8.5$ kpc. Radial
and vertical oscillations may be considered small
departures from an otherwise circular orbit (e.g. the vertical
oscillation has a period $P \simeq 62$ Myr and amplitude $\sim 35$ pc
\cite{BIN87}, smaller than the scale heights of the perturber
populations). Severe gravitational encounters are unlikely to have
distorted this orbit over the past 0.5 Gyr (i.e. two dynamical times)
making it reasonable to assume that the Sun has preserved its nearly
circular motion, with angular speed $\Omega_{\odot} \simeq 27$ km
s$^{-1}$ kpc$^{-1}$ ($v_{\odot} \simeq 230$ km s$^{-1}$). 

The spiral density waves trail the disk rotation with a characteristic pattern
speed $\Omega_p \simeq 19\pm5 $ k m s$^{-1}$ kpc$^{-1}$ \cite{WADA},
implying that the solar system streams through the arms at a mean
relative speed $v_r \simeq 68$ km s$^{-1}$. 
Due to the inherent difficulty of the measurement, the pattern speed
is not an accurately determined quantity and contributes 
the largest uncertainties to any estimate of the past structure of
the Galaxy.  Methods for 
estimating $\Omega_p$ have included measurement of the age gradient 
of objects along the Sagittarius-Carina arm (Avedisova 1989) and the velocity field of Cepheids (Mishurov et al. 1979).  Amaral and L\'epine (1997) 
estimated $\Omega_p$ based on a study of open clusters, an ideal population 
for such a study since their ages are well determined from the HR diagram. 
Their analysis suggests that $\Omega_p \simeq 20\pm5$ km s$^{-1}$kpc$^{-1}$. 
The estimate derived by Wada et al. (1994) is for the pattern speed of a 
putative end-on Galactic bar based on modeling of the molecular cloud 
longitude-velocity diagram.  Other arguments have been summarized by 
Amaral (1995) in favor of $\Omega_p \simeq 20~$km s$^{-1}$ kpc$^{-1}$.

The present day positions of the spiral arms have been outlined using
optical and radio observations of large H II regions (Georgelin \&
Georgelin 1976), supplemented by data from the 21-cm line of neutral
hydrogen, the H109$\alpha$ radio recombination line, and the 2.6-mm
line of carbon monoxide (which traces molecular hydrogen), used to
resolve distance ambiguities.  The highly excited H II regions define
two pairs of arms (four major arms altogether), which intersect the
solar orbit at angles of $10^{\circ}$--$12^{\circ}$. The face-on
morphology of the Galaxy is shown in Figure 1, along with the
location of the Sun during each of the six Phanerozoic extinctions.
The data for the major arm (the Sagittarius-Carina arm) and the
intermediate arm (the Scutum-Crux arm) are complete out to the solar
galactocentric radius of 8.5 kpc.  However, data is unavailable for
the internal arm (the Norma arm) beyond a galactocentric radius of
$\simeq 6.0$ kpc, due to obscuration by the Galactic center. We extend
this arm to the solar orbital radius using a logarithmic spiral model
\cite{AVE96}; this function provides excellent corroborating fits to
arms for which data do exist.

Figure 2 displays the times of solar spiral arm crossings in graphical
form (assuming a relative speed of 68 km s$^{-1}$) through the
Galactic free-electron distribution, as modeled by Taylor and Cordes
\cite{TAY93}, based on the radio and optical data of giant H II
regions and corroborated by $\gamma$-ray observations of Al-26, a
tracer of massive star nucleosynthesis \cite{CHE96}.  The free
electron density (or equivalently the ionized gas density) is a tracer
of spiral structure; ionized gas in the Galaxy is concentrated in the
H II regions surrounding hot OB stars, young star clusters, and in the
near exteriors and interiors of expanding supernova remnants.  Dotted
lines indicate the range of uncertainty in the past positions of the
spiral arms due to unwinding. (A simple way to estimate the unwinding
is to notice that the phase winding between the innermost regions of
the arms (the so-called inner Lindblad resonance where $\Omega_d
\approx 0$ km s$^{-1}$ kpc$^{-1}$ and $R \simeq 4.0$ kpc) to those at
radius $R_0$ is $\simeq \pi$ rad (Figure 1), over a time roughly equal
to the age of the Galactic disk $\simeq 12$ Gyr. The phase unwinding
for individual crossings is then $\sim 1^\circ$, $\sim 4^\circ$ and
$\sim 8^\circ$, respectively, for the first three spiral arms.  A more
detailed calculation gives $2.4^\circ$, $10^\circ$ and
$20^\circ$; Binney \& Tremaine 1987).

Along with the crossing times, Figure 2 illustrates the extinction 
timeline adopted from Sepkoski (1994), where individual
extinctions are modeled as gaussians after the subtraction of a mean
background extinction rate at each epoch. Notice a correlation between
the two timeseries, which {\it a priori} represent two rather
disparate temporal sequences, i.e., the solar spiral arm crossing
times and the geological times of terrestrial
extinction. Perhaps most interesting, as it involves the smallest extrapolation
in time, is the close coincidence of the K/T (Maastrichtian) event 
with the Sagittarius-Carina arm crossing 60 Myr-ago (0.6
\% the disk age) for the above-mentioned kinematic parameters.  Further
back in time, the end-Permian and upper Norian events coincide with
the crossing of the Scutum-Crux arm.  The upper Botomian
and possibly the late Ordovician (Ashgillian) extinctions may be associated with
the Norma arm, but the large uncertainty in the position of this arm
makes any definite association specious at best.  The late Devonian
(Frasnian) extinction does not coincide with any major arm, although a
detailed statistical examination \cite{HUB} of the extinction record
suggests that the Devonian and Norian events should only be regarded
as candidates for extinctions.  It is well worth mentioning that independent of
our assumed distance scale and kinematic model, the ratios of
timescales in the geologic record are a good match to those of the three spiral
crossings (assuming an unperturbed solar orbit).

The uncertainties involved in the above analysis are admittedly quite
large, and any quantitative comparison should naturally be regarded with 
caution.  Yet in the face of the terrestrial geologic record, and the near
certainty that at least one of the mass extinctions is linked to non-terrestrial
mechanisms, we find an idea which unifies these mechanisms appealing.  If a fraction of mass 
extinctions are indeed due to non-terrestrial phenomena, then the concentration 
of supernovae and other perturbers ought to make spiral arms far more hazardous than
other locations in the Galaxy.  A comparison between the extinction
record and the best available data on the spiral structure of the
Galaxy suggests that this may indeed be the case; although the large
uncertainties in the spiral arm pattern speed, as well as the
locations of the arms themselves prevent a more definitive comparison,
it is nonetheless intriguing that the observed spacing of the major
extinctions is approximately reproduced.  Moreover, evidence from the
fossil record \cite{OFF} (and possibly the Ir evidence; Yabushita \& Allen 1997) that the
mass extinctions were far more gradual than previously thought, lends
credence to the spiral arm hypothesis, as the Sun spends tens of Myr
in the vicinity of each arm, during which any or all of the
aforementioned processes are not only possible, but likely.  If both
boloidal impacts (the Iridium evidence) and supernovae are established
as culprits-in-common (as per the suggestions of Ellis, Fields and
Schramm 1996) from either geological or ice-layer records,
then spiral arm crossing must play an important role in the repeated
extinction of terrestrial (or extra-terrestrial) life.

\noindent
{\bf Acknowledgments:} 
We thank J. H. Taylor and J. M. Cordes for making their Galactic 
free-electron density
model widely available, as this paper makes use of their
software. We thank T. Padmanabhan, A. C. S. Readhead, M. R. Metzger and S. R. Kulkarni for several useful discussions.

\begin{figure}[ht]
\plotone{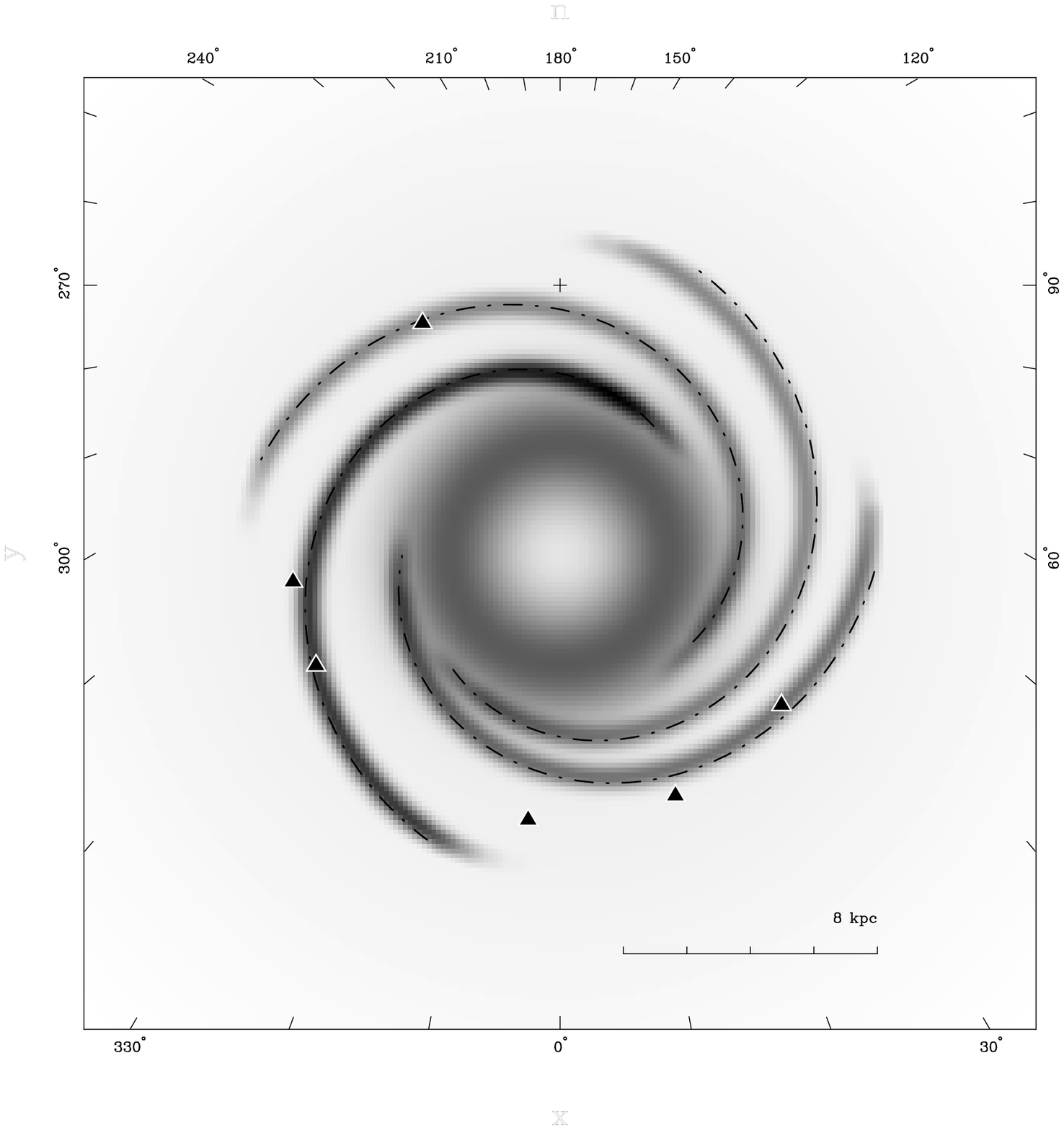}
\caption{\footnotesize Face-on view of the Galaxy.  Arms are fits of 
logarithmic spirals to the Taylor-Cordes model, extrapolated to $R\sim12$~kpc,
 with peak densities equal to the value at the solar radius.  Counterclockwise
 from the Sun (cross at top) are the Sagittarius-Carina, Scutum-Crux, 
Norma and Perseus arms.  Triangles mark the times of the major terrestrial extinctions.  The effect of unwinding
 is indicated by the dot-dashed lines defining the centroids of the arms for 
an unwinding of $1^\circ$, $4^\circ$ and $8^\circ$ for the first three arms, 
respectively.}
\end{figure}

\begin{figure}[ht]
\plotone{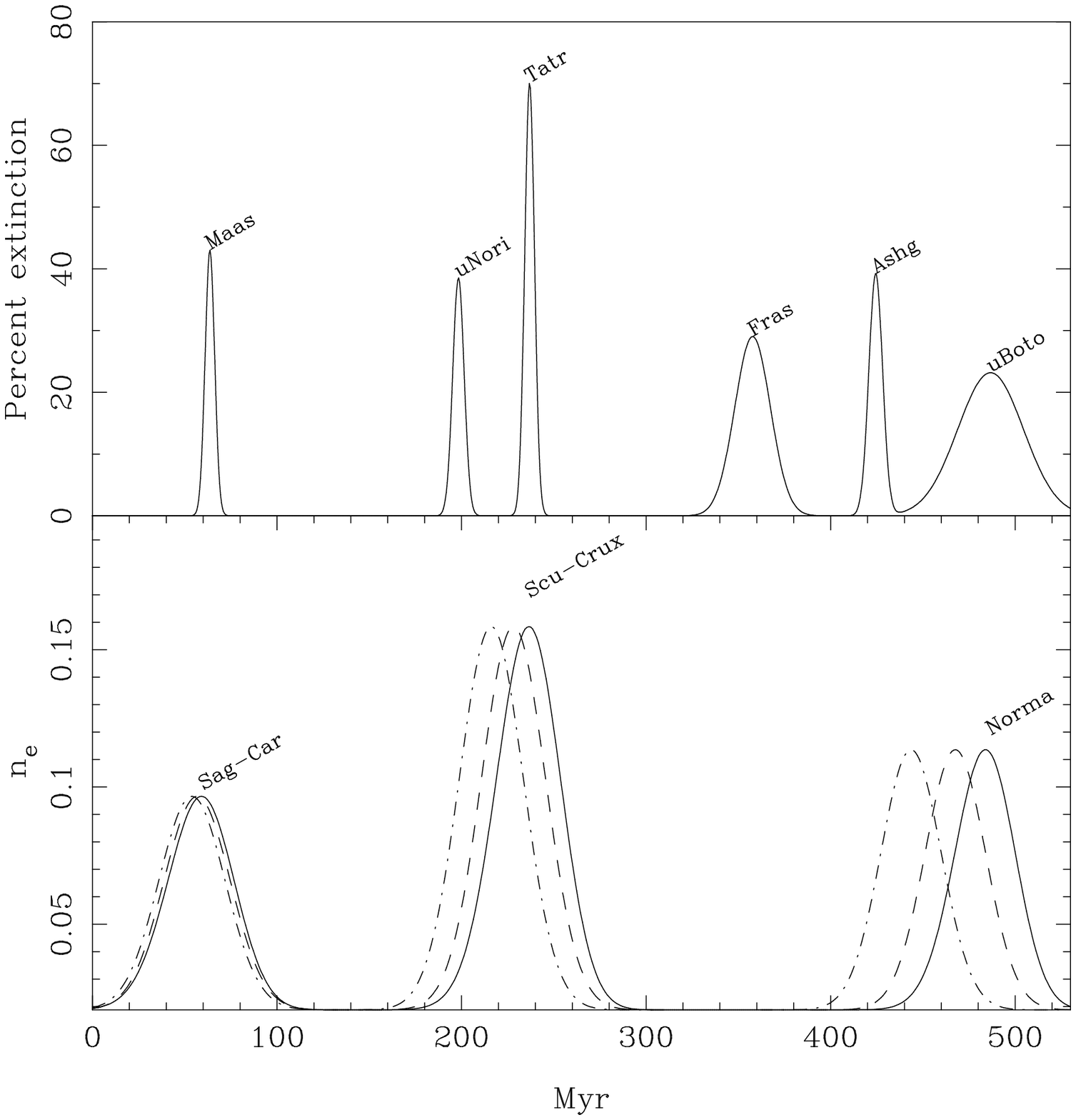}
\caption{\footnotesize (Top panel) Gaussian fits to peak percent extinctions 
(after Sepkoski \cite{SEP94}), after subtraction of a mean extinction level at each epoch. 
(Bottom panel) Spiral arm crossings (traced by electron density); locations 
and approximate widths of the spiral arms are taken from the model of Taylor 
and Cordes \cite{TAY93}.  The Norma arm is an extrapolation to the solar galactocentric 
radius of a fit to the Taylor-Cordes model.  The effect of unwinding is shown 
by the dashed line ($1^\circ$, $4^\circ$ and $8^\circ$ for the Sagittarius-
Carina, Scutum-Crux and Norma arms, respectively) and the dot-dashed line 
($2^\circ$, $10^\circ$ and $20^\circ$).}
\end{figure}

\section{APPENDIX}

In this appendix, we present a somewhat more detailed comparison of the two
timeseries shown in Figure 2.  Mass extinctions are identified as in
Sepkoski (1994).
As described in the text, the individual
extinctions are modeled as gaussians after subtraction of a mean
extinction rate at each epoch, estimated from the troughs in the
extinction record.  
We find that the cross-correlation of the two time series peaks at $\Delta t = 0$ for 
a relative velocity of $v_s = 68.4~$km/s (the expected value for
$\Omega_p \approx 19$ km s$^{-1}$ kpc$^{-1}$).

The statistical significance of the observed cross-correlation 
is assessed using Monte Carlo methods.  The locations of the individual extinctions
are uniformly randomized within the time bounds of the Phanerozoic 
period.  Two preconditions are applied when generating the
fake datasets: (i) the individual extinctions are not
allowed to overlap each-other within $2\sigma$-bounds (since
geological methods have to distinguish them as distinct
events), and (ii) wherever model extinctions do partially overlap,
their sum is truncated at the 100 percent level.
In addition, we let the orbital speed 
of the Sun in the frame corotating with the arms (i.e. $v_s = (\Omega_p
- \Omega_\odot)R_0$), be a gaussian random variable; that is, we let
$\Omega_p = 19\pm5$ km s$^{-1}$ kpc$^{-1}$ (1$\sigma$).
We generate $10^5$
fake extinction datasets, and for each we compute the zero-lag
cross-correlation with the spiral arm crossing curve.  We find that 99 percent
 of the randomly generated correlations were smaller than the actual data
correlation. The highest
tail-end cross-correlations, when examined carefully, displayed rough
positional interchange and strong clumping of the
extinction gaussians around the spiral arms, as expected.


\newpage

\footnotesize
\begin{thebibliography}{99}

\bibitem[]{AMA1} Amaral, L. H. 1995, PhD Thesis, Univ. Sa\~o Paulo, Instituto
Astron\^omico e Geof\'isico
\bibitem[]{AMA2} Amaral, L. H. \& L\'epine, J. R. D. 1997, MNRAS, 286, 885.
\bibitem[(Alvarez et al. 1990)]{ALV90} Alvarez, W.,  Asaro, F., \& Montanari, A., 1990, Science, 250, 1700-1702.
\bibitem[(Alvarez \& Muller 1984)]{ALV84} Alvarez, W. \& Muller, R. A., 1984, Nature, 308, 718-720.
\bibitem[(Avedisova 1989)]{AVE89} Avedisova, V. S. 1989, Astrophys., Vol 30, No. 1, 83.
\bibitem[(Avedisova 1996)]{AVE96} Avedisova, V. S., 1996, Astronomy Letters, 22, 443-454: Translated from Pisma v Astronomicheskii Zournal, 1996, 22, No., 7-8.
\bibitem[(Bignami \& Caraveo 1996)]{BIG96} Bignami, G. F. \& Caraveo, P. A., 1996, ARAA, 34, 331-381.
\bibitem[(Binney \& Tremaine 1987)]{BIN87} Binney, J. \& Tremaine, S., 1987, Galactic Dynamics, Princeton Univ. Press, 350.
\bibitem[(Chen et al. 1996)]{CHE96} Chen, W., Gehrels, N., Diehl, R., \& Hartmann, D., A\&AS, 1996, 120, 315-316,.
\bibitem[Clube \& Napier 1996]{CLU96} Clube, S. V. M. \& Napier, W. M., 1996, QJRAS, 37, 617-642.
\bibitem[(Crutzen \& Br\"uhl 1996)]{CRU96} Crutzen, P. J. \& Br\"uhl, C., 1996, PNAS, 93, 1582-1584.
\bibitem[Ellis \& Schramm 1995]{SCH95} Ellis, J. \& Schramm, D. N., 1995, PNAS, 92, 235-238.
\bibitem[(1996)]{SCH96} Ellis, J., Fields, B. D., Schramm, D. N., 1996, ApJ, 470, 1227-1236.
\bibitem[(1976)]{GEO76} Georgelin, Y. M. \& Georgelin, Y. P., 1976, A\&A, 49, 57-59.
\bibitem[(1970)]{HAT81} Hatfield, C. B. \& Camp, M. J., 1970, Bull. geol. Soc. Am.,81, 911-914.
\bibitem[(Hubbard \& Gilinsky 1993)]{HUB} Hubbard, A. E. \& Gilinsky, N. L. 1993, 1992, Paleobiology, 18, 148-160.
\bibitem[Koyama et al. 1995]{KOY95} Koyama, K., Petre, R., Gotthelf, E. V., Hwang, U., Matsuura, M., Ozaki, M. \& Holt, S. S., 1995, Nature, 378, 255-258.
\bibitem[(McCrea 1975)]{McC75} McCrea, W. H., 1975, Nature, 255, 607-609.
\bibitem[(Mishurov et al 1979)]{MISH} Mishurov, Y. N., Pavlovskaya, E. D., Suchkov, A. A. 1979, AZh, 56, 286.
\bibitem[(Officer et al 1987)]{OFF} Officer, C. B., Hallam, A. D., Drake, C. L. \& Devine, J. D., 1987, Nature, 326, 143-149.
\bibitem[Rampino \& Stothers 1984]{RAMP84} Rampino, M. R. \& Stothers, R. B., 1984, Nature, 308, 709-711.
\bibitem[Rampino \& Haggerty 1996]{RAMP96} Rampino, M. R. \& Haggerty, B. M., 1996, Earth Moon \& Planets, 72, 441-460.
\bibitem[(Raup \& Sepkoski 1982)]{RAU82} Raup, D. M. \& Sepkoski, J. J. Jr, 1982, Science, 25,1501-1503.
\bibitem[(Raup 1991)]{RAU91} Raup, D. M., 1991, Paleobiology, 17, 37-48.
\bibitem[(Ruderman 1974)]{RUD74} Ruderman, M. A., 1974, Science, 184, 1079-1081.
\bibitem[(1994)]{SEP94} Sepkoski, J. J., 1994, Geotimes, March 1994, 15-1.
\bibitem[(1984)]{SCH84} Schwartz, R. D. \& James, P. B., 1984, Nature, 308, 712-713.
\bibitem[(1993)]{TAY93} Taylor, J. H. \& Cordes, J. M., 1993, ApJ, 411, 674-684.
\bibitem[Thorsett (1995)]{THO95} Thorsett, S. E., ApJ, 1995, 444, L53-L55.
\bibitem[(van den Bergh \& Tamman 1991)]{VAN91} van den Bergh, S. \& Tammann, G. 1991, 1991, ARAA, 29, 363-407.
\bibitem[(Wada et al. 1994)]{WADA} Wada, K., Taniguchi, Y., Habe, A. and Hasegawa, T., 1994, ApJ, 437, L123.
\bibitem[(Yabushita \& Allen 1997)]{YABU} Yabushita, S. \& Allen, A., 1997, Astronomy \& Geophysics, 38 (2), 15.
\end{thebibliography}

\normalsize
\end{document}